\documentclass[pre,twocolumn,showpacs,preprintnumbers,amsmath,amssymb,dvips]{revtex4}
\usepackage[T1]{fontenc}
\usepackage[latin1]{inputenc}
\usepackage{times}
\usepackage{graphicx}
\usepackage{dcolumn}
\usepackage{bm}
\begin{document}
\title{Modeling self-organization of communication and topology in social networks}
\author{M. Rosvall$^{1,2}$}\email{rosvall@tp.umu.se}
\author{K. Sneppen$^{2}$}\homepage{http://cmol.nbi.dk}
\affiliation{$^{1)}$Department of Theoretical Physics, Ume{\aa} University,
901 87 Ume{\aa}, Sweden\\
$^{2)}$Niels Bohr Institute, Blegdamsvej 17, Dk 2100, Copenhagen, Denmark\\
}
\date{\today}

\begin{abstract}
This paper introduces a model of self-organization between communication
and topology in social networks, with a feedback between different
communication habits and the topology.
To study this feedback, we let agents communicate to 
build a perception of a network
and use this information to create strategic links.
We observe a narrow distribution of links when the
communication is low and a system with a broad distribution
of links when the communication is high.
We also analyze the outcome of chatting, cheating, and lying,
as strategies to get better access to information in the network.
Chatting, although only adopted by a few agents,
gives a global gain in the system.
Contrary, a global loss is inevitable in a system with too many liars.
\end{abstract}

\maketitle

\subsection*{Introduction}
Who communicates with whom and the social structure of a society are
strongly entangled. The social network reflects the
access to information that different parts of the system experience,
and social mobility may be seen as a quest for better information access.
A reliable global perception of the network,
often achieved by informal communication with acquaintances \citep{krackhardt,bass,brown,knoke,rogers,huberman,infoflow,rosvall},
makes the social mobility meaningful \citep{friedkin,dodds}.
The small talk consists in its simplest form of identifying
who to get the information from, and whom to transfer it to \citep{milgram,milgram1969,kleinberg,kleinberg-hierarchy,kleinberg-algorithm,wattsidentity}.
To understand the feedback between different communication habits and the topology,
we in this paper introduce an agent-based model that self-organize the social network.
That is, we allow agents to create new links to get easier access to some
parts of the system, based on interest and the information they obtained through 
communication with already established acquaintances \citep{interestrewire,friedkin-infoflow,trusina2004}.

After defining the model in the next section, we show that organized structures,
that can make use of the small-world properties of the network \citep{watts,kochen,perkins},
emerge when the communication is sufficiently high.
This is followed by an investigation of consequences of manipulating information.
What are the gains or costs when the agents adopt individual
strategies to get better access to the system on, respectively, local and global level?
We investigate consequences of chatting, cheating and lying,
and find, for example, that lying opens for a 
communication analogue to the prisoners dilemma game \citep{axelrod}.
Finally we explore a few variants of the model and, for example,
show how separation of interests naturally leads to
modular networks in the model.

\subsection*{Model}
Let us now define the model in detail, formulated in the two basic events:

\begin{itemize}
\item
\emph{Communication:}
Select a random link and let the two agents that it connects communicate about a 
random third agent \citep{bergmann}. The two agents also update their information about each other.

\item
\emph{Rewiring:}
Select a random agent and let it use its information to
form a link to shorten its distance to a randomly chosen other agent.
Subsequently a random agent loses one of its links.
\end{itemize}

\begin{figure}[tbp]
\centering
\leavevmode
\begin{tabular}{c}
\includegraphics[width=1.0\columnwidth]{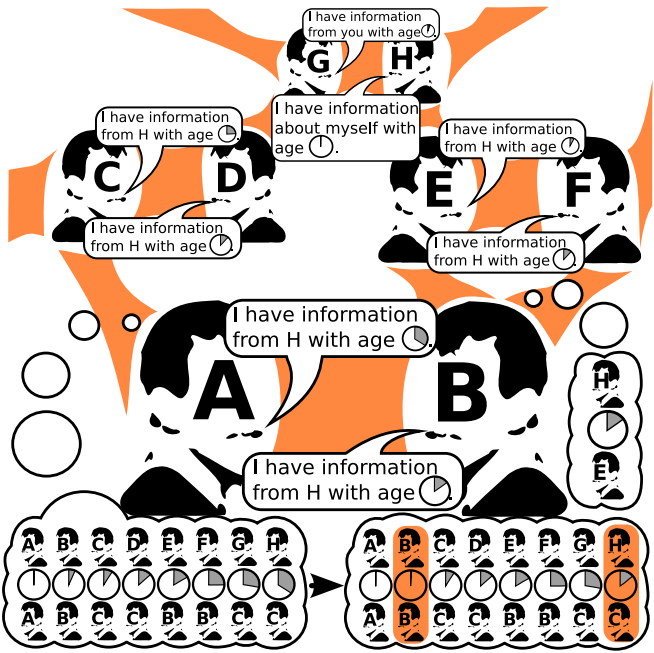}
\end{tabular}
\caption{ \label{fig1}(color online) \emph{Communication} as modeled in this
paper: Agents communicate with their neighbors in the
network about a third target agent, and estimate the quality of the
information by its age. The agent with the oldest information
adopts the viewpoint of the agent with the newest information.
Here, agent \textbf{A} learns that \textbf{B} has newer information about
\textbf{H}, disregards its old information,
and change its pointer towards \textbf{H} to \textbf{B}.
The information in the bottom bubbles are \textbf{A}'s knowledge
about the network based on communication with its neighbors
before and after the communication event with \textbf{B}:
For each agent (top row) the time of the most recent information is stored (middle row) together
with the acquaintance that provided the information (bottom row).}
\end{figure}

\begin{figure}[tbp]
\centering
\leavevmode
\includegraphics[width=1.0\columnwidth]{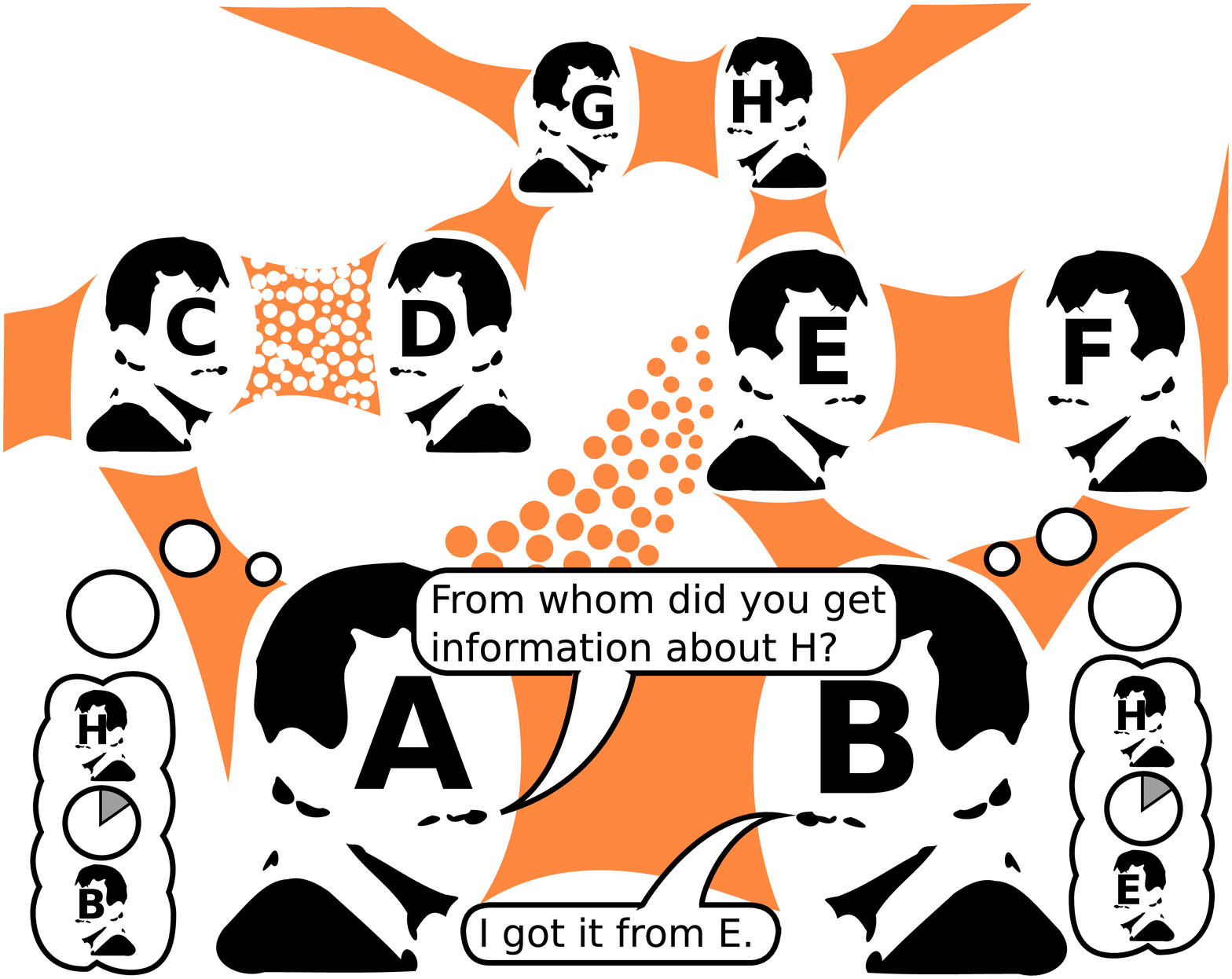}
\caption{\label{fig2}(color online) \emph{Rewiring} as modeled in this paper:
Agents create new contacts based on their available information (see Fig.\ \ref{fig1}).
In this example \textbf{A} uses \textbf{B},
the agent that \textbf{A} got the newest information about \textbf{H} from,
to get a better position relative to \textbf{H}.
\textbf{A} then creates a link to \textbf{E}, the agent that \textbf{B}
got it information about \textbf{H} from.
Thereafter a random agent looses a random link, in this example 
the connection between \textbf{C} and \textbf{D}.
}
\end{figure}

The communication event is typically repeated of the order of
number of links in the system for each rewiring event.
Figure \ref{fig1} and \ref{fig2} illustrate the two elements in the model.
The basic variables in the network model are nodes represented by $N$ agents
and $L$ links that correspond to the available communication channels in the system.
We let the agents communicate and in that way build their own perception of
where they are relative to other agents in the network.
Each agent, in Fig.\ \ref{fig1} exemplified by agent \textbf{A},
has a list of previously obtained information with entries for
each agent $i=$\textbf{A}, \textbf{B}, \textbf{C},$\dots$ 
For each entry $i$, the agent has a pointer to the 
agent that provided the most recent information
about $i$. This pointer is updated if someone else comes
with newer information about $i$ \citep{perkins}.
Therefore we also keep the age of all obtained information in
\textbf{A}'s memory (see clocks in Fig.\ \ref{fig1}).
The age of an agent's information about itself is always 0.
The age of any other information increases proportional to
the number of ongoing communication events in the system.
When two agents communicate about a third agent the agent with the older information 
disregards this and adopts the viewpoint of the agent with the newer information by
copying the age and changing the pointer.
In Fig.\ \ref{fig1} agent \textbf{A} communicates with \textbf{B} about agent \textbf{H},
and adopts the viewpoint of \textbf{B} because \textbf{B}'s information about \textbf{H} is newer.
\textbf{A} sets its clock for \textbf{H} to the same time as \textbf{B},
and change its pointer for \textbf{H} to \textbf{B}.
The age of the information serves as a qualifier that allows two communicating agents
to estimate which of them that have the most reliable information.

Figure \ref{fig2} describes the second main feature of the model,
the social mobility. We implement the social constraints of who 
can connect to whom by only allowing new links from an agent 
to acquaintances of its acquaintances \citep{bornholdt}.
A randomly chosen agent, here \textbf{A}, is interested in shortening its distance to another
randomly chosen agent in the system, here \textbf{H}. \textbf{A} therefore asks \textbf{B}, the agent that 
provided \textbf{A} with the newest information about \textbf{H}, about where the information came from.
\textbf{B} answers \textbf{E} and \textbf{A} builds a link to \textbf{E}
(if there is no link between \textbf{A} and \textbf{B}, \textbf{A} builds a link to \textbf{B}
and stops after that).
The creation of new links is balanced by random removal of links.
This is illustrated in Fig.\ \ref{fig2}, where \textbf{C}, chosen randomly,
looses its connection to \textbf{D}.

\begin{figure}[htbp]
\centering
\leavevmode
\includegraphics[width=1.0\columnwidth]{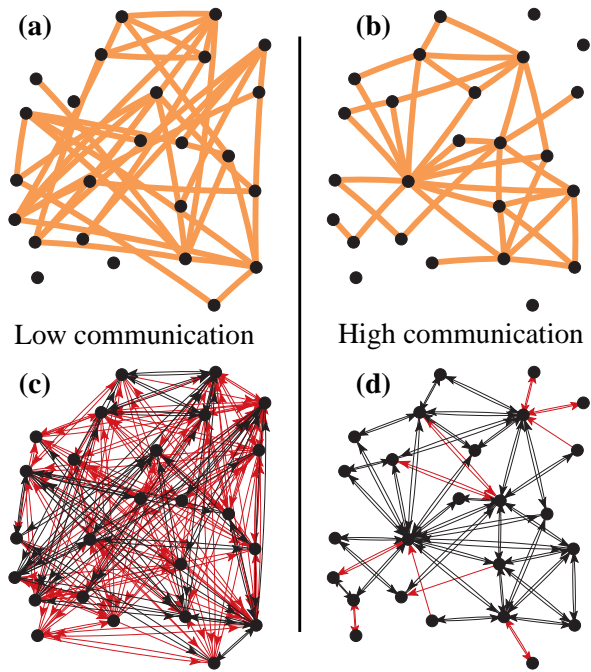}
\caption{ \label{fig3}(color online) Illustration of the to types of social bonds in 
the network at two different levels of communication. (a) and (b)
show the communication backbone over which agents communicate.
(c) and (d) show the perception network associated to the agents' directions
to other agents in the network.
The pointers are colored black when they are updated and coincide with active connections.
In the network with high communication (right panel), almost all pointers overlap with
the communication backbone.
}
\end{figure}

In the model, we thus have an interplay between 
the communication backbone network
and the perception that the agents have of this network.
The pointers of all agents, with both real and outdated links,
form the perception network.
In Fig.\ \ref{fig3} we illustrate the concept of a communication
backbone and the perception network at low and high communication
in a small network with 25 nodes and 38 links.
For relatively few communication events per rewiring 
(much less than the number of links in the system),
the communication and perception network diverge
and the rewiring that the agents perform has little to
do with the real topology of the network. As a consequence,
any rewiring of the network will be random and the network's overall topology
disorganize into a structure with a narrow degree distribution \citep{erdos}
(see the two networks to the right in Fig.\ \ref{fig3}).
In contrast, a high communication implies that new links are
introduced as a direct function of the present topology.
They are typically directed towards highly connected nodes since 
they provide new information.
With a tendency of building new links toward the majority
of the system, a reliable perception opens for positive feedback
and self-organization into a network
with broad degree distribution (see the two networks to
the right in Fig.\ \ref{fig3}).

\subsection*{Results}
\begin{figure}[!hbp]
\centering
\leavevmode
\includegraphics[width=1.0\columnwidth]{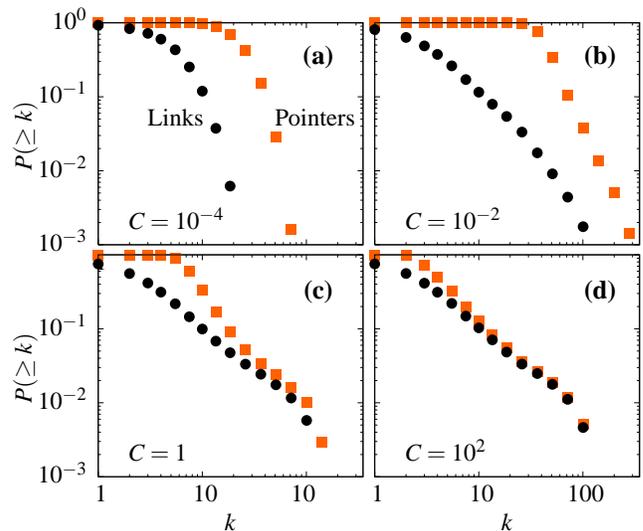}
\caption{ \label{fig4}(color online) Illustration of the feedback of communication on the topology
of both communication backbone and perception network at 4 different levels of communication $C$.
$C=1$ corresponds to on average 1 communication event per link and rewiring event.
Networks size is $N=1000$ agents connected by $L=2500$ links in the communication backbone.
}
\end{figure}

\begin{figure}[!hbp]
\centering
\leavevmode
\includegraphics[width=1.0\columnwidth]{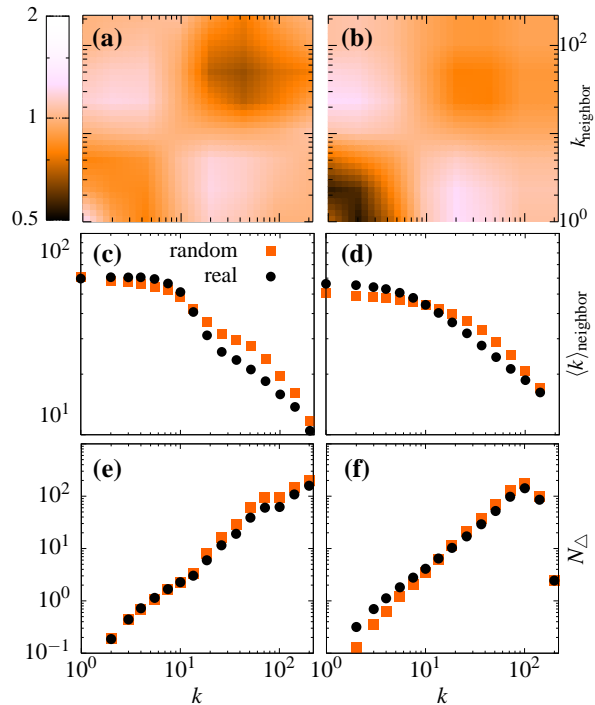}
\caption{ \label{fig5}(color online) The topology of the generated networks at two different
levels of communication $C$. $C=0.01$ in left panel and $C=1$ in right panel. First
row shows the correlation profile, second row the average neighbor degree as a function 
of degree and the third row the clustering measured as the number triangles.
All measures are compared with randomized counterparts of the networks with
unchanged degree-sequence.
}
\end{figure}

To quantify the interplay between the self-organization of network topology and
the overall communication level we in Fig.\ \ref{fig4} and \ref{fig5} show 
degree distributions for simulations of a system with $N=1000$ agents, $L=2500$ links,
and different values of the communication level $C$.
$C \cdot L$ is the number of communication events per rewiring event in the network,
and the degree $k$ of a node is its number of links.
We have also simulated networks with different number of links and found similar
results with a tendency towards more
pronounced non-random features with fewer links.
In Figure \ref{fig4} the number of communication events per rewiring and
link is varied between $C=10^{-4}$ and $C=100$.
At low communication level, $C<1$, the perception network has many more links
than the backbone network. As $C$ approaches  $C\sim 1$
the perception network prunes its links whereas the backbone network 
develops nodes with high degrees. At even higher values of $C$
the two networks converge toward the same broad degree-distribution.

Beyond the degree distribution, we in Fig.\ \ref{fig5} show the
correlation profile (top), the average neighbor degree (middle) and the number of triangles
(bottom) as a function of degree for low (left) and high (right) communication.
In all cases we compare with a randomized network where the degree sequence
is identical to the model generated, but all other features are
reshuffled \citep{maslov2002}.
We chose $C=10^{-2}$ as the low and  $C=1$ as the high communication level.
The overrepresentation of links
between nodes of high and low degree gives
extended community structures.
Triangles are overrepresented around
low-degree nodes and underrepresented
around high-degree nodes \citep{watts}.

All the presented results until now are based on agents that all
are the same. At any time their social position will however be 
different, because their sequence of communication and rewirings 
is strongly influenced by the history of the system.
The presented model describes a social game where the aim is to be central,
and a winner is an agent with many connections
that provide short and reliable communication to other agents.
The fact that we observe agents with a wide range of degrees
reflects the diversity of the possible outcomes of the game,
and raises the questions about whether there are some particular
strategies with which agents can improve their standing 
in the network? Can acting like a winner make you more likely to 
become a winner? Are there some particular situations where agents
systematically can attract additional connections and become a hub?

A highly connected agent became highly connected because it 
attracted new links by providing new information about other agents.
To provide new information is essential to win the game.
We therefore investigate a number of individual strategies
where agents attempt to convince other agents about their attractiveness
as an acquaintance.

\emph{Chatting} represents an increased communication rate. We let
the \emph{chatters} communicate twice as much as other agents by increasing
the probability that their links are chosen for a communication event by a factor 2.
Note that this also affects their acquaintances because they share links with 
the chatters.

\emph{Cheating} represents a decreased clock-speed. We let the \emph{cheaters}
use clocks that run at half the speed of the other agents' clocks, and
their information will thereby have a slower aging.
In practice they cheat by pretending that they have newer information than they really have.
Cheating might be either deterministic (every time unit is half length) or stochastic
(a time unit is counted with probability $1/2$).

\emph{Lying} represents a pure lie about the age of the information in a communication
event. Instead of updating the clock, the \emph{liars} replace the time by a random number.
Here we choose the random number between 1 and 100 that represents the typical
age of information about an agent within the second nearest neighbor
radius in a system with 1000 agents.

\begin{figure}[!hbp]
\centering
\leavevmode
\includegraphics[width=1.0\columnwidth]{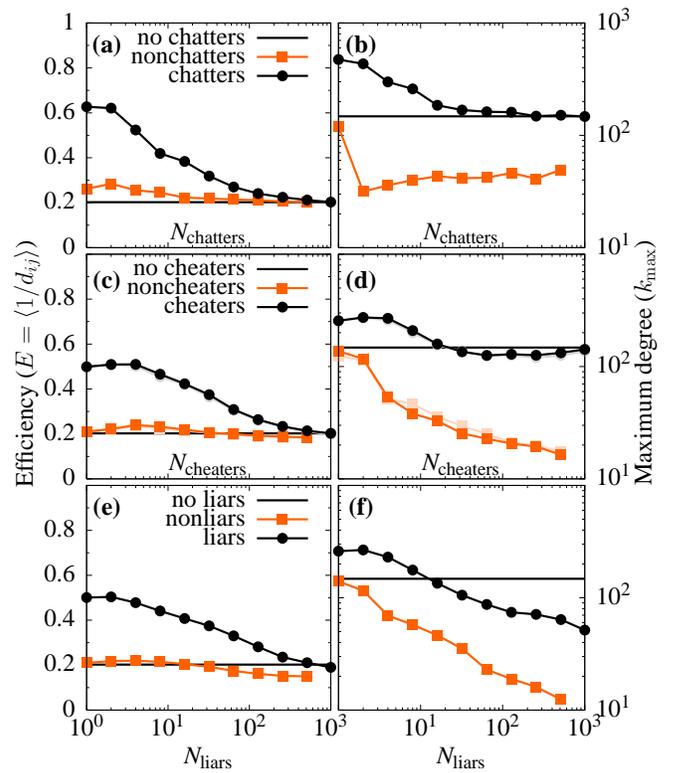}
\caption{ \label{fig6}(color online)%
Topological consequences of introducing strategic agents,
quantified through efficiency (defined as $E=\langle 1/d_{ij}\rangle$)
respectively maximum degree of both the strategic (black circles) 
and the remaining agents (orange squares).
(a-b)  Effect of having $N_{chatters}$ agents which communicate twice
as much as the remaining $N-N_{chatters}$ agents in network.
(c-d) Effect of $N_{\mathrm{cheaters}}$ agents that cheat
by running their internal clock at half the speed of the
other $N-N_{\mathrm{cheaters}}$  agents' clocks.
(e-f) Effect of a more brutal strategy where 
$N_{\mathrm{liars}}$ agents always pretend that their information about all other agents
is very new (of the order of what the remaining $N- N_{\mathrm{liars}}$ agents
have for their nearest or next nearest neighbors).
The communication rate is $C=1$ in a system
with $N=1000$ nodes and $L=2500$ links.
}
\end{figure}

\begin{figure}[!hbp]
\centering
\leavevmode
\includegraphics[width=1.0\columnwidth]{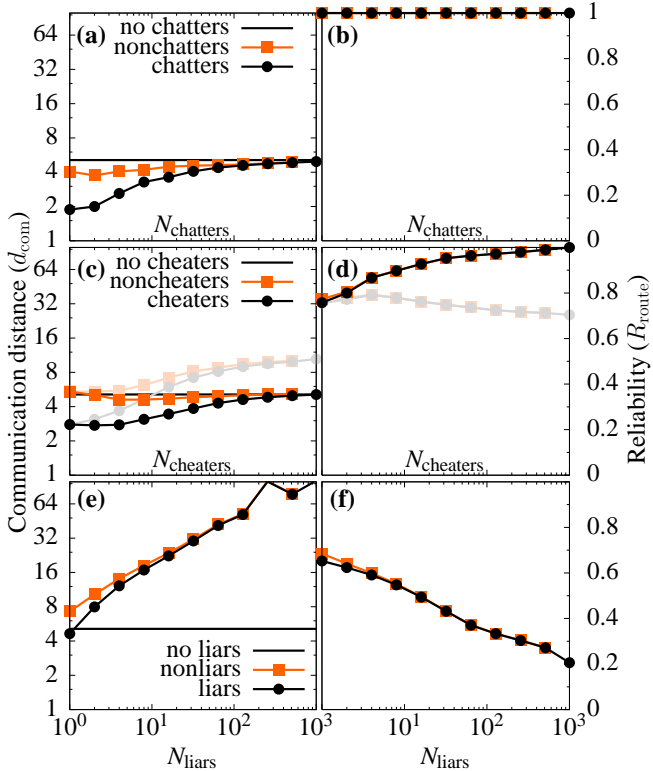}
\caption{\label{fig7}(color online)%
Perception consequences of introducing strategic agents,
quantified through communication distance, $d_{\mathrm{com}}$ and
reliability of routed messages, $R_{\mathrm{route}}$.
(a-b), (c-d), and (e-f), corresponds to the same strategies as in Fig.\ \ref{fig6}.
In (c-d) orange and black correspond to the deterministic cheaters and 
the faded colors correspond to the stochastic cheaters.
}
\end{figure}

In all three strategies the information is manipulated
to gain a local advantage.
However, there may also be a cost, both on local and on global scale.
This is what we examine in Fig.\ \ref{fig6} and \ref{fig7}.
Figure \ref{fig6} shows the topological consequences on the communication backbone and 
Fig.\ \ref{fig7} the effect on the perception network, as we vary the number of
strategic agents between 1 and the system size at communication level $C=1$.
The right panel in Fig.\ \ref{fig6} shows how the max degree of respectively the 
strategic agents (black circles) and non-strategic 
agents (orange squares) changes with $N_{\mathrm{strategic}}$.
When less than about 10 agents adopt any of the three strategies
the they gain in terms of degree.
However, as the number of liars increase, 
the overall network topology degenerates and it becomes impossible to 
sustain hubs. Also the liars become losers.
A more global examination of the effect of the various strategies are shown 
in left panel of Fig.\ \ref{fig6}.
The efficiency $E=\langle 1/d_{ij} \rangle$  
is the average value of the reciprocal distance \citep{latora} of, respectively,
the strategic agents, and the non-strategic agents.
This measure of typical distances in 
the network allow us to include also temporarely disconnected nodes.
In terms of efficiencies all strategies are successfully, 
and in addition they also seem to benefit the
other agents by providing short paths.

That the strategic agents become central in the communication backbone-network
does not directly imply that they can use their centrality.
The use of various strategies may influence the 
reliability of information that the agents have about the system
and thereby make long-distance communication more difficult.
In Fig.\ \ref{fig7} we examine the ability of agents to communicate 
across the system. 
$d_{\mathrm{com}}$ in the left panel is the average number of agents that
participate in communicating a message from an agent to another agent.
For chatters, we again see that everybody gains. For cheaters, on the other hand,
everybody gains if the cheating is deterministic, but already a few agents
with stochastic cheating (faded) makes communication across the system less efficient.
The from Fig.\ \ref{fig6} seemingly successful strategy of lying completely destroy
the communication abilities (Fig.\ \ref{fig7}(e)). 
One single liar makes some benefit of its strategy, but
two liars are enough to not only destroy for the nonliars,
but also for the liars themselves.

To emphasize this result, we in the right panel of Fig.\ \ref{fig7} show the
reliability, $R_{\mathrm{route}}$, of the perception network.
To calculate $R_{\mathrm{route}}$,
we send messages between any pairs of node and let the intermediate agents
route the messages with their pointers.
A message fails when it reaches an agent for the second time and 
the path forms a loop. $R_{\mathrm{route}}$ is the fraction 
of messages that reach the target.
The chatters are able to keep perfect reliability, but the cheaters and 
especially the liars destroy it. When there are 1000 deterministic cheaters the
reliability is again $100\%$ (see Fig.\ \ref{fig7}(d)). This is because it only corresponds to a rescaling of time
when all agents are deterministic cheaters.
The liars, examined in Fig.\ \ref{fig7}(f), 
systematically destroy the signaling capacity of the network.

The presented model is the simplest in a family of models
based on an interplay between communication and dynamical changes of topology.
We have investigated a range of variations,
including versions where each agent has a biased interest in other agents.
For example, we let an agent's target of interest be chosen inversely proportional
to the age of the information about the target \citep{java}.
Thereby interests are focused around the neighborhood and we observe an
increase in the number of triangles in the system.
In another variant, we divided the agents into several interest groups.
By increasing the probability to communicate and move inside
the interest group, the network develops a modular topology.


\subsection*{Discussion}
In this work we have introduced a model framework
that allow us to investigate the interplay
between social structures and communication habits.
We have shown that low communication
leads to random networks with narrow degree distributions.
Increased communication naturally gives
nonrandom structures characterized in particular
by social networks with broad degree distributions.
In addition to developing broad degree distributions, the
networks also tend to organize highly connected agents to connect
preferably to low connected agents.

With the model, we have investigated how manipulating information
influence the social structure, quantified by the topology 
of the emerging network.
Firstly we increased the communication frequency of individual agents.
The result was striking, the more an agent chats with its surroundings,
the better it performs.
Increased chatting requires increased effort, but
our model shows that there is both a local and a global gain to this effort.

Secondly we investigated the effect of cheating with the age of the 
information a particular agent distributes. If an agent only
underestimated the time since it received the information, the agent 
improved its position but at a cost to the remaining system.
As cheating does not cost more communication 
effort of the agent, it is the cheap way to optimize the social position selfishly.
However, already a single cheater decreases the overall reliability 
to send signals across the network, reflecting a moderate global 
cost to this strategy. 

Thirdly we investigated the more violent strategy of lying. The lying
agents pretend that they have recent information about everybody else.
The strategic agents in this way succeed to attract links and thereby become central in
the communication backbone network.
However, only a single liar in a system with non liars benefit from the strategy.
Lying is so destructive that one liar is enough to break down the
reliability of the network and none is in reality a winner.

\subsection*{Summary}
In a broad perspective the proposed model suggests an 
information theoretical perspective on social and possibly also 
economic systems. By introducing an information game based on social links 
and communication rules we present an
approach to the dynamics of human organization.
Agents in a network use information, obtained through
communication in the network, to form new links for better access to information.
The introduced feedback enables us to study the topological consequences
of different communication habits. 
By playing this communication game, we learn that communication,
although not equally distributed, is a benefit for everyone.
However, communication is expensive.
Other cheaper strategies are tempting,
but a strategy based on lies easily counteract the intention to have better
access to information. The social possibilities 
are not solely defined by the position in the network, but 
also by the quality of the surrounding information.

\end{document}